\documentclass[aps,prl,reprint,groupedaddress,showpacs]{revtex4-1}
\usepackage{graphicx}%
\usepackage{amsmath}%
\usepackage{color}

\begin{document}


\title{Ultrafast doublon dynamics in photo-excited 1\emph{T}-TaS$_2$}


\author{M. Ligges$^1$}
\email[]{manuel.ligges@uni-due.de}
\author{I. Avigo$^1$}
\author{D. Gole\v{z}$^2$}
\author{H. U. R. Strand$^2$}
\author{L. Stojchevska$^1$}
\author{M. Kall\"{a}ne$^3$}
\author{P. Zhou$^1$}
\author{K. Rossnagel$^3$}
\author{M. Eckstein$^4$}
\author{P. Werner$^2$}
\author{U. Bovensiepen$^1$}
\affiliation{$^1$Faculty of Physics, University of Duisburg-Essen, 47048 Duisburg, Germany}
\affiliation{$^2$Department of Physics, University of Fribourg, 1700 Fribourg, Switzerland}
\affiliation{$^3$Institute of Experimental and Applied Physics, University of Kiel, 24098 Kiel, Germany}
\affiliation{$^4$Max Planck Research Department for Structural Dynamics, University of Hamburg-CFEL, 22761 Hamburg, Germany}

\date{\today}

\begin{abstract}
Strongly correlated systems exhibit intriguing properties caused by intertwined microscopic interactions that are hard to disentangle in equilibrium. Employing non-equilibrium time-resolved photoemission spectroscopy on the quasi-two-dimensional transition-metal dichalcogenide 1\emph{T}-TaS$_2$, we identify a spectroscopic signature of double occupied sites (doublons) that reflects fundamental Mott physics.
Doublon-hole recombination is estimated to occur on time scales of one electronic hopping cycle $\hbar/J\approx$~14~fs.
Despite strong electron-phonon coupling the dynamics can be explained by purely electronic effects captured by the single band Hubbard model, where thermalization is fast in the small-gap regime. Qualitative agreement with the experimental results however requires the assumption of an intrinsic hole-doping. The sensitivity of the doublon dynamics on the doping level provides a way to control ultrafast processes in such strongly correlated materials.
\end{abstract}





\maketitle

Complex matter is characterized by strong interactions between different microscopic degrees of freedom, often resulting in rich phase diagrams where tiny variations of controllable parameters can lead to significant changes of the macroscopic material properties \cite{dagotto2005}. This competition or coexistence often occurs on comparable energy scales and, thus, is only partly accessible in the spectral domain. The dynamics of such systems, driven out of equilibrium by an external stimulus, can shed new light on the underlying short- and long-range interactions because different coupling mechanisms result in dynamics on experimentally distinguishable femto- to picosecond time scales \cite{gianetti2016}. In contrast to materials with well-defined quasi-particles, the theoretical description and analysis of nonequilibrium phenomena in strongly correlated electron systems is challenging \cite{Aoki14}. Model studies have predicted intriguing effects of electron-phonon couplings \cite{Eck13,Gol12,Wer15}, spin excitations \cite{Gol214,Kog14,Eck14} and dynamical screening \cite{Gol15},
but connecting these insights to measurements on real materials has rarely been attempted. In this Letter, we consider the photo-induced electron population dynamics in a quasi-two-dimensional system with strong electron-electron and phonon-phonon interactions and a finite density of defects. Our combined theoretical and experimental effort provides an example of how such competing processes can be disentangled in the time domain, and how insights from theory can help identify the nature of photo-excited carriers and the dominant relaxation pathway.

1\emph{T}-TaS$_2$ is a layered crystal that exhibits a manifold of electronically and structurally ordered phases \cite{wil75,sip08,Sto14}. In its high temperature state ($T>542$ K), the system is undistorted and metallic, while cooling results in the formation of various charge density waves (CDW) with an increasing degree of commensurability and a transition to semiconductor-like behavior. Below the critical temperature of 180~K, a commensurate periodic lattice distortion (PLD) is formed, giving rise to the formation of ``David-star"-shaped 13-Ta-atom cluster sites. This structural distortion is accompanied by a rearrangement of the partially filled Ta 5\emph{d} band into sub-manifolds. The uppermost half-filled band is prone to a Mott-Hubbard transition, forming an occupied lower Hubbard band (LHB) representing a single particle population per cluster and an unoccupied upper Hubbard band (UHB) indicating double population of cluster sites. On-site Coulomb interaction acts on these doublons and leads to an energy gap of a few 100 meV between the LHB and UHB \cite{dar92}. This widely accepted picture was recently challenged \cite{Rit15}. It was proposed that the formation of an energy gap can be explained by orbital texturing, which implies that Mott physics is not primarily responsible for the insulating state. Our combined theoretical and experimental effort shows how such ambiguity can be addressed under non-equilibrium conditions. We identify a hierarchy of time scales which allows for specific studies of electronic correlation effects in complex materials.

Time- and angle-resolved photoemission spectroscopy is a powerful tool for exploring the ultrafast electronic response of 1\emph{T}-TaS$_2$ on femto- to picosecond time scales \cite{per06,per08,pet11,hel12,avi16}. Our present studies were carried out in a pump-probe scheme in normal emission geometry on \emph{in situ} cleaved single crystals, which exhibit non-perfect stoichiometric ratios (1\emph{T}-Ta$_{(1-x)}$S$_2$ with $x$ $\approx$ 0.01) \cite{End00}. The samples were excited with 50 fs laser pulses from a regenerative Ti:Sa laser amplifier ($\hbar\omega_\text{pump}=$1.55 eV) and subsequently probed by direct photoemission using frequency-quadrupled pulses ($\hbar\omega_\text{probe}=$6.2 eV) \cite{avi16}.
The pump-probe cross-correlation was determined to be $110\pm5$ fs (Gaussian full width at half maximum) through the fastest observed response and the corresponding maximum was set as time zero.
The spectral resolution of 80 meV was determined by analyzing the width of the high-temperature Fermi edge (assumed to be rigid). Incident excitation fluences $F$ were kept well below the critical energy density necessary to drive the system thermally into the nearly-commensurate CDW phase \cite{hel10}.

\begin{figure}
\includegraphics [width=0.5\textwidth]{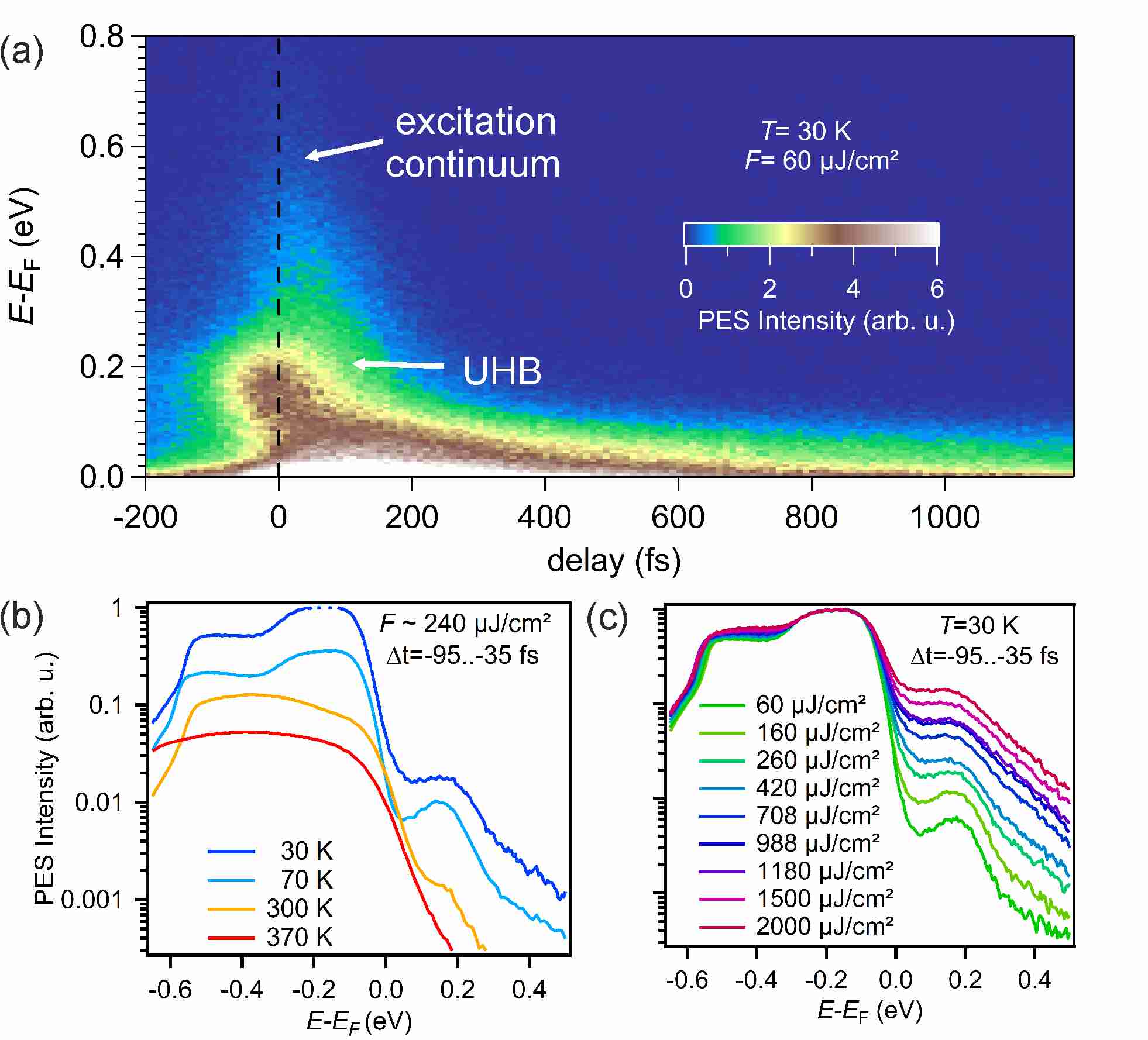}
\caption{
\label{fig:fig1}(color). (a) False-color representation of the time-dependent photoelectron intensity above $E_{\mathrm{F}}$ at 30 K for $F$=60 $\mu$J/cm$^2$ in normal emission. Beside an excitation continuum, the upper Hubbard band is observed at $E-E_{\mathrm{F}}\approx 175$ meV in the vicinity of $t=0$. (b) and (c) show photoemission spectra after excitation for selected $T$ and $F$, respectively. The spectra were averaged within $\Delta t$=-95 .. -35 fs. Curves in (b) are offset for better visibility.}
\end{figure}

Our discussion here focuses on the data obtained in the weak excitation limit and at a low base temperature of $T$=30 K, a situation in which we expect only minor modifications of the CDW-ordered state. Assuming that every absorbed pump photon excites one valence electron on one cluster site once, the excited electron density in the first atomic layers was estimated to be $<3\%$ for $F$=100 $\mu$J/cm$^2$ \footnote{This estimation is based on the optical properties reported in \cite{bae75} and a geometrical site density of $n=7\cdot10^{13}$cm$^{-2}$ \cite{Yam83}}. Under such excitation conditions we observe the appearance of a photoemission peak at $E-E_{\mathrm{F}}\approx 175$ meV that is barely visible for higher $F$ and $T$, even when the transient spectra are averaged over the relevant delay range (Fig.~1b-c). The contrast between the photoemission peak and underlying background differs between samples of different growth batches, a finding that we assign to the variation of stochiometric composition and, thus, to different hole doping. The general behavior, however, is the same for all samples under investigation. Fig.~1a shows a false-color representation of time-dependent photoemission spectra. Upon pumping, a broad excitation continuum is generated that reaches up to $E-E_{\mathrm{F}}\approx1.5$ eV. The population decay of this continuum is resolved and the closer to $E_{\mathrm{F}}$ the intensity is analyzed, the longer the relaxation times become -- a behavior well known for electronic excitations at metal surfaces \cite{Bau15}. In contrast, the sharp spectral signature at 175~meV responds significantly faster. Due to its fast dynamics (Fig.~2), we interpret this feature as the UHB that directly reflects double occupation of cluster sites. This assignment is corroborated by the observed energy of the feature, which is in agreement with recent scanning tunneling microscopy studies \cite{cho15}. It is furthermore observed that the UHB intensity gradually decreases with increasing temperature (Fig.~1b), which is consistent with the emergence of a macroscopic coexistence of insulating and conducting domains (corresponding to the different CDW states) up to the critical temperature of 542 K, where the full transition to the normal metallic phase occurs \cite{sip08}.

The distinct response of the UHB becomes evident when the transient energy distribution curves are considered (Fig.~2a). The UHB spectral weight reaches its maximum around $t=0$ and is lost after 100~fs. This loss is accompanied by ultrafast filling of the gapped region around $E-E_{\mathrm{F}}\approx -0.2 \ldots 0.2$ eV. At later delays, only intensity of the broad continuum remains.

\begin{figure}
\includegraphics[width=0.5\textwidth]{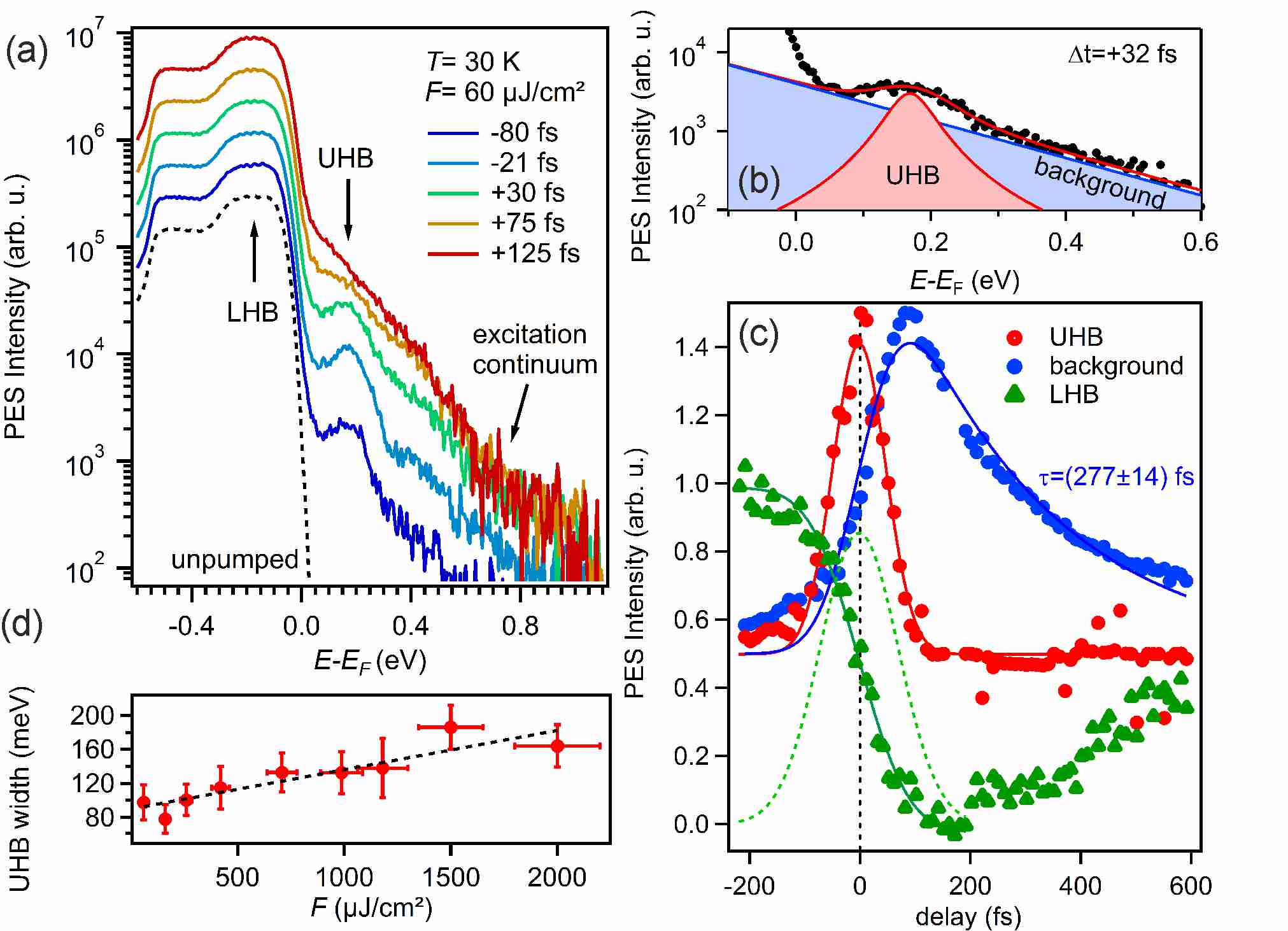}%
\caption{\label{fig:fig2}(color). (a) Transient photoemission spectra obtained for $T$=30 K and $F$=60 $\mu$J/cm$^2$. The curves are offset for better visibility. (b) Exemplary fit to the transient energy distribution curve obtained at $\Delta t=$+32 fs. The superposition of the continuum and the Lorentzian line was convoluted with the instrumental energy resolution function. (c) Temporal evolution of the UHB spectral signature in direct comparison to the underlying continuum and electronic gap, as well as the LHB intensity loss. The temporal derivative $-\frac{dI}{dt}$ of the LHB dynamics fit is shown as a dashed line for direct comparsion with the UHB signal. Note that all curves are rescaled for better visibility. (d) Spectral width of the UHB signature as a function of excitation fluence $F$.}
\end{figure}

In order to further discuss the temporal evolution of the UHB and separate its dynamics from the underlying continuum, we decomposed both spectral contributions by fitting the energy distribution curves with an exponential background and a Lorentzian line, as shown exemplarily in Fig.~2b. The fit results are shown in Fig.~2c and reveal the ultrafast response of the UHB in contrast to the slower dynamics of the spectral background that exhibits a population decay time of 277~fs. This background is still present at positive delays, when the UHB intensity has disappeared. Correspondingly, the UHB is not populated by secondary excitations and we conclude that the photo-induced dynamics can not be described by incoherent scattering processes considering rigid bands. The temporal evolution of the UHB intensity is the fastest response observed in our experiments and we set it´s maximum to time zero. Analyzing the dynamics (Fig.~2c), we estimated the time scale of doublon-hole recombination to be less than 20 fs due to the absence of an exponential decay component within experimental uncertainties. Remarkably, this time scale matches approximately the electronic hopping cycle $\hbar/J\approx$~14~fs estimated by dynamic mean field theory \cite{per08}. This is confirmed by the fact that the temporal evolution of the UHB correlates with the decrease
of the LHB signature that was previously shown to occur quasi-instantaneously on a time scale faster than 20 fs \cite{pet11} and serves as an additional timing signal. However, this correlation does not necessarily imply a direct optical transition between the LHB and the UHB, whose energy difference ($\sim$400 meV) is largely exceeded by the pump photon energy (1.55 eV). We rather consider a simultaneous depopulation of the LHB by excitation into higher lying bands and a population of the UHB due to an excitation from lower lying states, as prominent features in a similar energy range were also observed in the optical properties of 1\emph{T}-TaS$_2$ and assigned to transitions from the p-like valence band \cite{bae75}. This suggests the importance of not only static, but also photo-induced holes for the UHB dynamics in our experiment. Under the excitation conditions used here, the LHB intensity is reduced by only 5 \% and it can be expected that the charge ordered state is only weakly perturbed.

For a given excitation fluence we find that the UHB line profile is independent of $t$ (data not shown), indicating that the population within the UHB thermalizes on a time scale that can not be resolved in our experiments. We observe that the energetic width of the UHB signature increases linearly when higher excitation fluences are used (Fig.~2d), excluding the photoemission line to arise from an unoccupied rigid band populated by the 6 eV probe pulse. We also note that the energy of the UHB line is independent of $t$, which
indicates that the signature is not related to
polaronic excitations discussed in the literature \cite{dea11}. Our finding rather demonstrates the ultrafast decoupling of electronic and lattice degrees of freedom, since polarons would result in an energetic stabilization on phononic time scales. Indeed, the response of the UHB is faster than a quarter-period of the highest-frequency phonons in 1\emph{T}-TaS$_2$ of 11.9~THz \cite{Gas02}, which also implies a decoupling of the doublon dynamics from the periodic lattice distortion associated with CDW formation.

Our experimental findings are partly incompatible with the recently proposed energy gap formation based on orbital texturing \cite{Rit15}. Optical excitation of an orbitally ordered system could in principle result in the loss of order due to increased entropy and scattering. The fastest time scale we can imagine is determined by electron-electron scattering which leads to the relaxation dynamics of the excitation continuum, see Fig. 2a. However, the observed loss of the UHB occurs earlier than that, a fact explained in the present study by local electron-electron correlations. We thus conclude that, while orbital order as well as layer stacking might play a role in electronic band formation, a discussion of the experimentally observed ultrafast electronic response in close vicinity to $E_F$ requires a scenario where electron-electron correlation effects are important.

\begin{figure}
\includegraphics[width=0.5\textwidth]{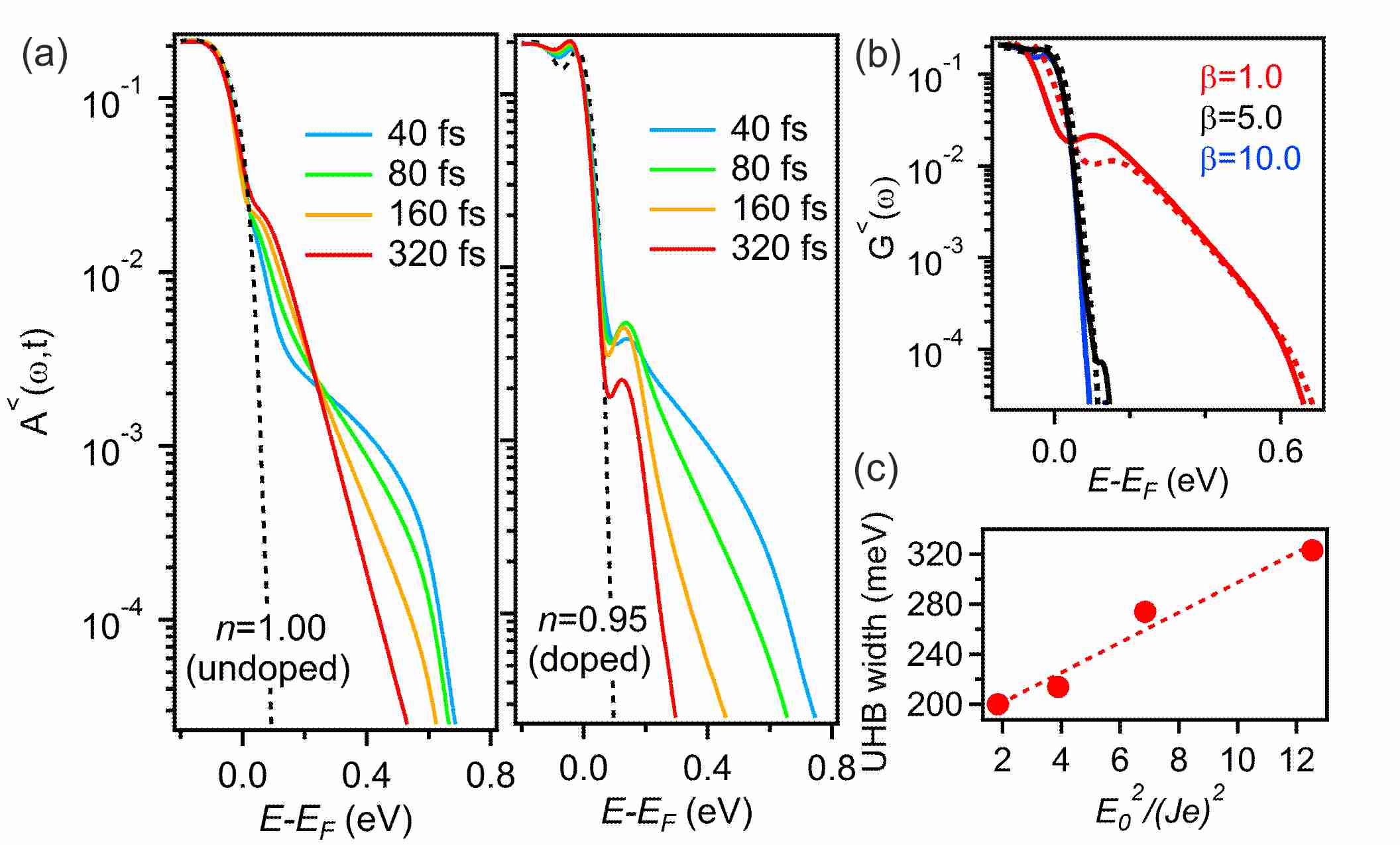}
\caption{\label{fig:theory}(color). (a) Time evolution of the occupation function $G^<(\omega,t)$ for a half-filled band ($n$=1, left panel) and in the hole-doped case (n=0.95, right panel). The dashed lines show the equilibrium situation before excitation. (b) Equilibrium occupation function $G^<(\omega)$ for $n$=0.98 (solid lines) and $n$=0.95 (dashed lines) at different temperatures $\beta=(k_bT)^{-1}$, given in units of $W$ ($\beta=1$ corresponds to a temperature of 32300 K). (c) Spectral width of the UHB signature at $E-E_{\mathrm{F}}\approx 175$ meV as a function of excitation density $E_0^2$. The dashed line indicates a linear dependence.}
\end{figure}

In the following we discuss a simple theoretical picture based on a purely electronic model and demonstrate how strong interaction is crucial for describing the ultrafast dynamics. We simulate the dynamics of a single band Hubbard model on a two-dimensional triangular lattice
\begin{multline}
H=\sum_{i\delta\sigma}J c_{i+\delta,\sigma}^{\dag}c_{i+\delta,\sigma}+\mu n_{i} 
+U\sum_{i}(n_{i\uparrow}-\frac{1}{2})(n_{i\downarrow}-\frac{1}{2}),
\end{multline}
where $c_{i\sigma}^{\dag}$ denotes the creation operators for a Fermion on lattice site $i$ with spin $\sigma$, $J$ is the hopping integral between neighboring sites, $\mu$ is the chemical potential, $n_{i}$ is the number of carriers on site $i$ and $U$ is the on-site Coulomb repulsion. The electric field of the pump laser $E(t)$ is applied along the (1,1) direction and is incorporated via the Peierls substitution. The parameters were chosen such that, in the absence of an external perturbation ($E(t)=0$), this Hamiltonian mimics the equilibrium spectral function of single-layer 1\emph{T}-TaS$_2$ \cite{per06,cho15} with a bandwidth of $W$=0.36 eV.

The dynamics after optical excitation is modeled by perturbing the system with a Gaussian pulse of the form $E(t)=E_{0}\exp(-4.6(t-t_{0})^{2}/t_{0}^{2})\sin(\omega(t-t_{0}))$, where the duration of the pulse $t_0$ is chosen such that it accommodates a single optical cycle. The frequency $\omega$ was chosen to be $\omega/J=8.0$ and the pulse amplitude was $E_0/(J e)=2.0$. While this pulse frequency generates direct transitions between LHB and UHB, we have also checked the scenario of photo-doping from lower-lying bands by temporarily coupling an occupied (empty) electron bath to the UHB (LHB). The relaxation dynamics for both protocols is consistent, since it is mainly governed by the excess kinetic energy of the excited doublons. The amplitude of the excitation does not alter the qualitative dynamics as long as one restricts to the weak excitation regime. To solve the electron dynamics we use the nonequilibrium dynamical mean field theory (tDMFT) \cite{Aoki14}, which maps a correlated lattice problem onto a self-consistently determined impurity problem \cite{georges1996}. To treat the impurity problem we use the lowest order strong coupling expansion, the noncrossing approximation (NCA). To confirm that the resulting dynamics is qualitatively correct, and not sensitive to the details of the bandstructure, we also employed the one-crossing approximation (OCA) on the Bethe lattice. Realistic gap sizes are obtained for $U=0.36$ eV in NCA, and $U=0.43$ eV in OCA.

The transient occupation dynamics is analyzed in terms of the partial Fourier transform of the lesser component of the Green's function $A^{<}(t,\omega)=\text{Im}[\int_t^{t+t_{max}} dt' e^{i \omega (t'-t)}G^{<}(t',t)]$ (occupied denity of states). We will first discuss the response of the ideal Mott insulator at half band filling ($n$=1), see Fig. 3a.
The optical excitation leads to a partial occupation of the UHB that relaxes within the band and results in a slow population build-up at the lower edge. This prediction of a monotonously increasing UHB population is inconsistent with the experimental observation. It is, however, in agreement with previous studies \cite{werner2014} which show that the thermalization in an isolated small-gap insulator can lead to an increase in double occupation on the timescale of few inverse hoppings. We note that in contrast to the previous theoretical interpretations \cite{per06,per08}, but in agreement with the experimental data (and recent arguments based on high temperature expansions \cite{perepelitsky2016}), our simulations do not predict a substantial gap filling after photoexcitation.

A more realistic description is obtained if one considers an effectively hole-doped system 1\emph{T}-Ta$_{(1-x)}$S$_2$. Assuming that every Ta atom out of a 13 atom superstructure cluster donates four electrons to the surrounding S atoms, we estimate that the effective filling of the sub-band straddling $E_F$ can range from half filled ($x$=0, $n$=1) to quarter-filled ($x$=0.01, $n$=0.5). As the exact stochiometric composition of our samples is hard to determine with sufficient accuracy, we assume a small doping level of $n$=0.95 to discuss the general influence on the UHB dynamics (Fig. 3b). In contrast to the half-filled case the occupation function in the doped case shows a transient increase of the doublon spectral weight, which quickly vanishes. This evolution is in qualitative agreement with the experimental finding.

In agreement with previous works \cite{eckstein2011,werner2014} the Hubbard band in the small gap regime thermalizes on the time scale of several inverse hoppings, which we confirmed by checking that the fluctuation-dissipation theorem is fulfilled. Therefore we can compare the results in the long-time limit with the thermal states at elevated temperatures, see Fig.~\ref{fig:theory}(b). In the half-filled case, the UHB of this small gap system is always substantially occupied due to the finite overlap of the high-temperature Fermi-Dirac distribution function with the UHB. In the doped case, the Fermi-Dirac distribution is shifted to lower energies and the overlap with the UHB is exponentially suppressed. In such a situation, a significant population in the UHB can only be achieved at extremely high electron temperatures. This also explains the experimental requirement of low excitation densities to observe the ultrafast reduction of the UHB population and is reflected in the experimental finding that the spectral width of the UHB increases linearly with excitation fluence (Fig.~2d), which is also reproduced in the theoretical calculation (Fig.~3c). We note that even though electron-phonon coupling is important in 1\emph{T}-TaS$_2$, we can neglect the phonon dynamics on electronic time scales.
In any event, the effect of electron-phonon interactions, as well as short-ranged spin excitations, would be to speed up the relaxation and thermalization \cite{WerEck15,EckWer16}. This likely explains the faster dynamics in the experiment, compared to the simulations which neglect this physics.

In summary, we identified in a combined experimental and theoretical study the double electron population of cluster sites in 1\emph{T}-TaS$_2$ and estimated their relaxation time to be on the order of the electronic hopping cycle $\hbar/J$. Essential was the time domain approach, which facilitated the detection of doublons before further excitations like secondary electrons or phonons set in. We emphasize the importance of static or photoexcited holes which mediate the ultrafast relaxation on hopping timescales. This tunability of femtosecond dynamics in close energetic vicinity of the Fermi level may enable novel design principles for, e.g. ultrafast solid state switches. More generally, the ability to identify the spectral function, including its non-equilibrium dynamics as a signature of strong electron correlations, will lead to improved microscopic understanding of complex materials with competing interactions.

We acknowledge financial support by the Deutsche Forschungsgemeinschaft through SFB 616, SPP 1458, SFB 1242 and FOR1700 and from ERC Starting Grant No. 278023. L.S. acknowledges the Alexander von Humboldt Foundation. We also thank R. Sch\"{u}tzhold and S. Biermann for fruitful discussions.

\bibliography{TaS2}

\end{document}